\documentclass{mn2e}
\usepackage{url} 
\usepackage{amsmath}
\usepackage{amssymb}
\usepackage{graphicx}
\usepackage{epsfig}
\usepackage{ulem}
\usepackage{color}
\usepackage{IEEEtrantools}

\def\tilde{\widetilde}
\def\spose#1{\hbox to 0pt{#1\hss}}
\def\lta{\mathrel{\spose{\lower 3pt\hbox{$\mathchar"218$}}
        \raise 2.0pt\hbox{$\mathchar"13C$}}}
\def\gta{\mathrel{\spose{\lower 3pt\hbox{$\mathchar"218$}}
        \raise 2.0pt\hbox{$\mathchar"13E$}}}

\newcommand{\be}{\begin{equation}}
\newcommand{\ee}{\end{equation}}
\newcommand{\apj}{ApJ}
\newcommand{\apjl}{ApJ}

\newcommand{\aap}{A$\&$A}

\newcommand{\mnras}{MNRAS}

\newcommand{\prd}{PRD}

\newcommand{\nat}{Nature}

\newcommand{\apss}{Astrophysics \& Space Science}
\newcommand{\physrep}{Physics Reports}

\begin{document}

\title[Tilt angle distribution in newly born NSs]
{The distribution of tilt angles in newly born NSs: role of interior viscosity and magnetic field}
\author[]{Simone Dall'Osso$^1$, Rosalba Perna$^1$\\
$^1$ Department of Physics and Astronomy, Stony Brook University, Stony Brook, NY, USA}
\maketitle

\begin{abstract}
We study how the viscosity of neutron star (NS) matter affects the
distribution of tilt angles ($\chi$) between the spin and magnetic
axes in young pulsars. Under the hypothesis that the NS shape is
determined by the magnetically-induced deformation, and that the
toroidal component of the internal magnetic field exceeds the poloidal
one, we show that the dissipation of precessional motions by bulk
viscosity can naturally produce a bi-modal distribution of tilt
angles, as observed in radio/$\gamma$-ray pulsars, with a low
probability of achieving $\chi \sim (20^\circ - 70^\circ)$ if the
interior B-field is $\sim (10^{11} - 10^{15})$~G and the birth spin
period is $\sim 10 - 300$~ms. As a corollary of the model, the idea
that the NS shape is solely determined by the poloidal magnetic field, or by
the centrifugal deformation of the crust, is found to be inconsistent
with the tilt angle distribution in young pulsars. When applied to
the Crab pulsar, with $\chi \sim 45^\circ - 70^\circ$ and birth spin
$\gtrsim$ 20 ms, our model implies that: (i) the magnetically-induced
ellipticity is $\epsilon_B \gtrsim 3 \times 10^{-6}$; (ii) the
measured {\it positive} $\dot{\chi} \sim 3.6 \times 10^{-12}$ rad
s$^{-1}$ requires an additional viscous process, acting on a timescale
$\lesssim 10^4$ yrs. We interpret the latter as crust-core coupling
via mutual friction in the superfluid NS interior. One critical
implication of our model is a GW signal at (twice) the spin frequency
of the NS, due to $\epsilon_B \sim 10^{-6}$. This could be detectable
by Advanced LIGO/Virgo operating at design sensitivity.
\end{abstract}
\section{Introduction}
The interior structure of neutron stars (NS) can affect their
rotational dynamics in measurable ways. Discontinuous exchanges of
angular momentum between normal matter and the superfluid components
in the crust and/or core can lead to sudden timing irregularities,
{\it i.e.} glitches.  Anisotropic stresses in the NS crust/core cause small deviations from sphericity in the NS shape, typically
measured in terms of the ellipticity $\epsilon$ (e.g. Alpar \& Pines 1985). In turn,
such deviations induce a precessional motion if they are not
perfectly aligned with the NS spin axis. The precessional motion is also 
sensitive to the dynamics of the superfluid interior, and to its coupling to the 
normal matter (e.g., Shaham 1977, Alpar \& Sauls 1988, Sedrakian et al. 1999, Link 2003, 2006, 
Andersson et al. 2006).

The inertia associated to the dipole magnetic field provides a minimum 
``effective" ellipticity for a magnetized NS, $\epsilon_{\rm min} \sim 10^{-13} B^2_{12}$ 
(Zanazzi \& Lai 2015). 
Crustal deformations are limited by the maximum breaking strain of the
crustal lattice, implying an ellipticity $\epsilon_{c} <
10^{-6}$ (Horowitz \& Kadau 2009): however, for isolated NS, realistic sources of 
strain likely produce much smaller values of $\epsilon_c < 10^{-8}$, unless
special circumstances occur (e.g. Jones 2012 and references
therein). A major source of anisotropic stress in the core is the
NS magnetic field, which produces an ellipticity\footnote{We will
  write $Q_n$ for a quantity $Q$ in units of $10^n$.} $\epsilon_B \sim
10^{-10} B_{13}^2$ for normal NS matter, or a factor $\sim 100$ times
larger when protons in the core are superconducting (Baym et al. 1969, 
Easson \& Pethick 1977, Cutler 2002).
As such, magnetic deformations may even dominate a NS ellipticity
provided that the core magnetic field is $\gtrsim 10^{13}$ G.

NS precession is hard to detect due to its small amplitude and long
characteristic timescale (typically $\sim$ yrs). However, due to its
diagnostic potential, it has been searched for decades: to date, its
detection has been claimed in a couple of objects (Stairs et al. 2000,
Kramer et al. 2006), plus a handful of additional candidates (Lyne et
al. 2010), showing periodic modulations in timing properties well
correlated with pulse profile changes. Their natural interpretation in
terms of freebody precession has been challenged by detailed studies, 
that revealed a complex pattern in the periodic modulations not easily related to simple precession (Lyne et
al. 2010). However Jones (2012) and, more extensively, Ashton et
al. (2016) have revived the case for a precession interpretation of
these objects.

Precessional motion might also be damped by viscosity in NS interiors,
in particular early in a NS life, with possible implications for the
distribution of tilt angles $\chi$, {\it i.e.} the angle between the
spin and magnetic axes (Jones 1976).  For a biaxial ellipsoid,
depending on whether it is {\it oblate} or {\it prolate}, viscous
dissipation will cause the spin and symmetry axes to become aligned or
orthogonal, respectively (Mestel \& Takhar 1972). Consequently, the
tilt angle $\chi$ will also decay or grow on the viscous timescale,
much shorter than the spindown time in young NS (Jones 1976).

Observationally, pulsar tilt angles have been studied by various
authors. Tauris \& Manchester (1998) first noted a preference for tilt
angles to be either $\lesssim 40^\circ$ or $\gtrsim 80-90^\circ$, with
fewer objects at intermediate values. They also found hints of
alignment of the magnetic and spin axes, with an estimated timescale
$\tau_{\rm align} \sim 10^7$ yrs. The latter conclusion, with a
somewhat shorter timescale, was reached by Young et al. (2010), while
Rookyard et al. (2015a) found a similar bi-modality in the tilt angle
distribution in a sample of young, gamma-ray emitting radio
pulsars. The apparent alignment might be consistent with the effect of
the electromagnetic torque (e.g., Goldreich 1970, Jones 1976). The
lack of pulsars with intermediate tilt angles, and the abundance of
small tilt angles even in pulsars with spindown ages $< \tau_{\rm
  align}$ (Rookyard et al. 2015a,b), are very hard to reconcile with
the hypothesis that they reflect a random tilt distribution at birth,
as frequently assumed in the literature. Alternatively, the magnetic
axis of the NS could be oriented close to the direction of its spin,
soon after formation. This might be the case if, e.g., the
  NS fast rotation had an important influence on the helicity of the
  birth magnetic field (Braithwaite \& Spruit 2004, Braithwaite \&
  Nordlund 2006), or if the NS magnetic field resulted from a dynamo
  in the proto-NS phase, during which differential rotation plays a key
  role (e.g., Braithwaite 2006).  As the observed distribution of
tilt angles is not consistent with either of these hypotheses, it
appears likely that it rather reflects some evolutionary process.  In
particular, the bi-modality seen in the young pulsars of Rookyard et
al. (2015a) suggests that, along with long-term alignment, some faster
process is also at work.

Recently, Lyne et al. (2013) measured an {\it increase} of the tilt
angle in the Crab pulsar, at the rate $\dot{\chi} \sim 0.62^{\circ}$
per century. The latter was shown to be possibly consistent with
freebody precession of the pulsar, for particular combinations of the
NS ellipticity and tilt angle (Philippov et al. 2014, Zanazzi \& Lai
2015). In this interpretation, the measured {\it positive}
$\dot{\chi}$ is a transient effect, associated with half of the
precession cycle, the secular trend being an alignment driven by the
magnetic dipole torque.  An alternative explanation for the positive
$\dot{\chi}$ of the Crab pulsar could be the dissipation of precession
energy, {\it if the NS shape is distorted into a prolate
  ellipsoid}. This requires that the magnetic field in NS interiors is
dominated by a toroidal component - such that precessional dynamics
causes $\chi$ to grow over time - and that the magnetically-induced
deformation dominates over crustal or other types of stress.

Motivated by these findings, we reconsider and expand
(Sec. \ref{sec:physical-mech}-\ref{sec:viscosity-NS}) the idea first
proposed by Jones (1976), that viscous dissipation of freebody
precession in newly born NS can produce large tilt angles at ages
$\lesssim 10^3$ yrs. With respect to previous discussions of the
subject we (i) update the microphysics description, including effects
of a realistic NS EoS and a detailed treatment of fluid motions in the
precessing NS core; (ii) consider the effect of shear viscosity at
late times, when the core temperature is $\lesssim 10^9$ K and protons are
superconducting; (iii) explore the implications of a wide range of
initial conditions on the final tilt angle distribution, and show that
a bi-modal distribution of tilt angles at early age - as is observed -
may be expected (Sec. \ref{sec:evolution}).

We then generalize our model to explain the measured positive value of
$\dot{\chi}$ in the Crab pulsar, and propose an interpretation
(Sec. \ref{sec:Crab}) in which viscous dissipation of precession
energy is provided, in this object, by crust-core coupling via mutual
friction. We then conclude (Sec. \ref{sec:conclusions}) that mutual
friction might affect, on longer timescales, the tilt angle
distribution in NS before alignment kicks in.

\section{General scenario}
\label{sec:physical-mech}
Our work is based on a number of observation- and
theory-driven assumptions, which we briefly summarize in the following in order to
clarify their validity and scope.
\subsection{Observed distribution of tilt angles}
\label{Sec:observational}
Our starting point is provided by the following observational facts and their interpretation:
\begin{itemize}
\item[(i)] The distribution of pulsar tilt angles is not consistent
  with a random distribution at birth (Tauris \& Manchester 1998,
  Rookyard et al. 2015a). While several caveats might affect the
  estimated tilt angles (Rookyard et al. 2015b), we assume that the
  overabundance of low-tilt pulsars, and the paucity of
  intermediate-tilt ones, are real effects.
\item[(ii)] In the sample of young (spindown age $< 10^6$ yrs),
  $\gamma$-ray emitting radio pulsars of Rookyard et al. (2015a),
  truly orthogonal rotators ($\chi > 80^\circ$) are, if anything,
  over-represented with respect to a flat distribution. The striking
  feature, in this relatively small sample, is the lack of pulsars
  with $40^\circ \lesssim \chi \lesssim 80^\circ$: based on this, it
  seems reasonable to consider the tilt angle distribution of at least
  these young pulsars as double-peaked, or bimodal.
\item[(iii)] Several authors have found hints of a long-term alignment
  in the pulsar population on a timescale $\sim 10^6 - 10^7$ yrs
  (Tauris \& Manchester 1998, Weltevrede \& Johnston 2008, Young et
  al. 2010). We assume that such effect is real, and that it can be
  accounted for by the electromagnetic torque.
\item[(iv)] Long-term alignment driven by the electromagnetic torque
  can explain, at least in part, the over-abundance of low tilt angles
  in the pulsar population. However, the fact that a similar effect is
  found in young radio pulsars, where orthogonal rotators are also
  relatively abundant, suggests that some other process might already
  be favoring either low or large tilt angles, on a significantly
  shorter timescale.
\end{itemize}
\subsection{Model assumptions}
\label{sec:theoretical}
Our proposed theoretical framework is summarized here:
\begin{itemize}
\item[(i)] The alignment time due to the electromagnetic torque is
  $\gtrsim 10^6$ yrs, if\footnote{The numerical value also depends on
  the NS magnetic dipole.} $P/\cos\chi \gtrsim 0.7$ ``at birth" (Jones
  1976). For plausible values of pulsar birth spins, this requires
  $\chi$ to be very close to $90^\circ$. Then, either all NS are born
  nearly orthogonal or their tilt angle grows rapidly to $\approx
  90^\circ$, before alignment kicks in. We will focus on the latter
  idea, originally proposed by Jones (1976).
\item[(ii)] The tilt angle in the Crab pulsar is indeed measured to
  increase at a rate of $\dot{\chi} \approx 0.62^\circ$/century (Lyne et
  al. 2013). This measurement could provide support to the above idea,
  although the interpretation of $\dot{\chi}$ is not unique (Sec. 6).
\item[(iii)] Tilt angles grow quickly to $\approx 90^\circ$ by viscous
  damping of precession in an oblique rotator\footnote{Having the
    magnetic axis tilted to the spin axis.} (Mestel \& Takhar 1972,
  Jones 1976). For this to work, the NS must have non-spherical shape
  with the largest axis of inertia (almost) aligned to the magnetic
  axis. The latter requires that (a) the deviation from sphericity is
  dominated by the magnetic field and (b) the magnetic field in the NS
  interior is predominantly toroidal, distorting the NS into a prolate
  ellipsoid (e.g., Cutler 2002, Dall'Osso et al. 2009). Note that, if
  (b) is not met, magnetic stresses will produce an {\it oblate}
  ellipsoid, in which case viscous dissipation drives $\chi$ towards
  {\it zero}.
\item[(iv)] A generic stable configuration for a NS magnetic field is
  that of a twisted-torus, in which a toroidal-poloidal B-field is
  contained in a torus-shaped region in the NS core, threaded by the
  large scale dipole (Braithwaite \& Nordlund 2006). Stability arguments suggest that 
  the toroidal component can exceed the poloidal one even by very large factors\footnote{The
    presence of a large toroidal field in the NS interior is invoked
    to explain the properties of magnetars (Thompson \& Duncan 2001,
    Perna \& Pons 2011, Dall'Osso et al. 2012). Here we are assuming a
    similar geometry, at lower field strength, for all NS.}  (e.g.,
  Reisenegger 2009, Braithwaite 2009, Akg\"{u}n et al. 2013), although, for barotropic 
  equations of state, it has been proven that the opposite may also be true 
  (Lander \& Jones 2009, Lander 2013).
  %while the opposite is notpossible.
\item[(v)] Tilt angles at birth might be very small, if the mechanism
  that amplifies the NS magnetic field inherits the direction of its
  spin. This seems plausible, if the magnetic field in the NS core has
  the twisted-torus shape discussed above. We will focus on this case,
  aiming at explaining the bi-modal distribution found in the young
  pulsar sample of Rookyard et al. (2015a). At the end of Sec. 5, in
  the light of our results, we will also discuss the possibility that
  the SN explosion produces a wider range of tilt angles at birth.
\item[(vi)] The newly formed NS is completely fluid as long as its
  temperature is $>T_{\rm cryst} \sim 4 \times 10^9$ K. At such
  temperatures, dissipation is dominated by bulk viscosity.  Below
  $T_{\rm cryst}$, the crust starts to form and new dissipative
  processes become possible: in this work, we will only consider bulk
  (and shear) viscosity in the NS fluid core, thus focusing on a
  ``minimal dissipation" scenario. Possible effects of the crust will
  be discussed in Sec. 6, in relation to the Crab pulsar.
\end{itemize}
\subsection{Main definitions}
\label{sec:definitions}

When the fluid NS undergoes precession, the absence of rigidity
doesn't allow its structure to sustain non-hydrostatic stresses. Thus,
a secondary flow is established inside the object (Mestel \&Takhar
1972, Lander \& Jones 2017), having the precession frequency \be
\label{eq:define-precession}
\omega \equiv \epsilon \Omega {\rm cos}\psi\, , \ee where $\epsilon$
is the ellipticity, $\Omega$ the
rotation frequency and $\psi$ the angle between the symmetry and spin axes. 
The timescale for dissipation of the freebody precession is \be
\label{eq:define-taud}
\tau_{\rm d} \equiv \frac{2 {\rm E}_{\rm pre}}{\left| \dot{{\rm E}}_{\rm diss} \right|} \, ,
\ee
where the  precession energy is 
\be 
\label{eq:precession-energy}
E_{\rm pre} = \frac{1}{2} I \epsilon \Omega^2 \cos^2\psi ~~~~~~{\rm
  for~a~prolate~ellipsoid} \, .  \ee Because of point (iii) in
Sec. \ref{sec:theoretical}, we only consider prolate ellipsoids from
now on, as they are required for $\chi$ to grow. When the magnetic
field dominates the NS deformation, the magnetic axis is almost
coincident with the symmetry axis: in this case, the tilt angle
$\chi~(\approx \psi)$ between the magnetic and spin axes can be
substituted in Eq. \ref{eq:precession-energy}. The dissipation rate,
$\dot{\rm E}_{\rm diss}$, depends on the type of viscosity and will be
defined below for each process considered.  Finally, the timescale for
the growth of the tilt angle $\chi$ is related to $\tau_{\rm d}$ as
\be
\label{eq:tauchi-def}
\tau_\chi \equiv \frac{{\rm sin} \chi}{\displaystyle \frac{d}{dt} {\rm
    sin} \chi} \, .  \ee Equating $\dot{{\rm E}}_{\rm pre}$
to $\dot{{\rm E}}_{\rm diss}$ in Eq. \ref{eq:define-taud}, we
therefore obtain \be \tau_\chi = \frac{{\rm sin}^2 \chi}{{\rm
    cos}^2\chi} \tau_{\rm d} \, \ee and, from this,  the
evolution equation for the tilt angle \be
\label{eq:evolution-equation}
\dot{\chi} = \frac{{\rm cos} \chi}{{\rm sin} \chi~\tau_{\rm d}}\, . \ee

\section{Viscosity of neutron star matter}
\label{sec:viscosity-NS}
In this section we discuss both bulk and shear viscosity in the NS
fluid core, deriving expressions for the associated energy dissipation
rate and the corresponding dissipation timescales
(cf. Eq. \ref{eq:define-taud}). The effects of NS cooling and the
role of baryon condensation will be discussed in
Sec. \ref{sec:supercond}.

In full generality, we can write the energy dissipation rate due to bulk
viscosity as (Friedmann \& Sergioulas 2013) \be
\label{eq:define-bulk-dissipation}
\dot{{\rm E}}^{({\rm bulk})} _{\rm diss} \equiv \int \zeta |\nabla \cdot
\mathbf{\delta v}|^2 = \omega^2 \int \zeta(\rho, T, x) \left|
\frac{\Delta \rho}{\rho}\right|^2 dV \, ,  \ee where
$\zeta$ is the bulk viscosity coefficient and $x$ the charged particle fraction.  
The second step derives from $\nabla \cdot \mathbf{\delta v} = i \omega \Delta
\rho/\rho$ (Lindblom \& Owen 2002), where $\Delta \rho$ is the
Lagrangian compression accompanying fluid motions: its maximum value
is obtained when $\Delta \rho \approx \delta \rho$, where $\delta
\rho$ is the non-spherical component of the density perturbation due
to the NS spin (Mestel \& Takhar 1972, Lander \& Jones 2017). Later (sec. \ref{sec:compressibility}) we will discuss a
general relation between $\Delta \rho$ and $\delta \rho$, identifying a regime in which 
they are approximately equal.

The corresponding expression for shear viscosity is 
\be
\label{eq:define-shear-dissipation}
\dot{{\rm E}}^{({\rm shear})} _{\rm diss} \equiv 2 \int \eta \delta
\sigma^{ab} \delta \sigma_{ab} ~dV \, , \ee 
where $\delta \sigma_{ab}
= \nabla_a \delta v_b + \nabla_b \delta v_a - \displaystyle
\frac{2}{3} \delta_{ab} \nabla_c \delta v^c$, and $\eta$ represents
the shear viscosity coefficient. To zeroth order, the ratio between
the two dissipation rates is mostly determined by the ratio between
$\zeta$ and $\eta$ (e.g., Cutler \& Lindblom 1987): this will be
discussed further in the next subsections.

\subsection{Bulk viscosity}
\label{sec:bulk-coefficient}
The coefficient $\zeta$ can be expressed in terms of fundamental
physical properties of the NS (e.g., Lindblom \& Owen 2002) \be
\label{eq:def-general}
\zeta \equiv \frac{ \tau_\beta ~n \displaystyle \frac{\partial
    p}{\partial x} \displaystyle \frac{dx}{dn}}{1+\left(\omega
  \tau_\beta \right)^2}\,, 
\ee 
where $n$ is the baryon number density, 
$p$ the pressure, \be
\label{eq:taubeta}
 \tau_\beta = \frac{6.9}{T^6_{10}} \left(\frac{\rho}{\rho_{\rm n}}\right)^{2/3} ~~{\rm s} \, 
 \ee
 is the $\beta$-reaction equilibrium timescale for pure $npe$ matter (Reisenegger \& Goldreich 1992), 
and $\rho_{\rm n} \approx 2.7 \times 10^{14}$ g cm$^{-3}$ is the nuclear saturation density. 
 \subsubsection{The $\beta$-equilibrium timescale}
 Expression (\ref{eq:taubeta}) for pure $npe$ matter neglects
 interactions among the baryons, that determine the NS EoS, 
 and considers only the neutron branch of modified Urca reactions. In a
 more realistic model of NS matter, three factors contribute to
 increase the $\beta$-reaction rate, thus decreasing 
 $\tau_\beta$ (cf.  Dall'Osso \& Stella 2017): {\it a)} the nuclear
 symmetry energy, $S_v(n)$, which describes baryon interactions 
 at supra-nuclear density; {\it b)} the appearence of more
 particles, e.g. muons at density $\gtrsim 2.2 \times 10^{14}$ g
 cm$^{-3}$, which adds new channels
 for Urca reactions; {\it c)} the proton branches of all modified Urca
 reactions, that provide a non-negligible 
 neutrino emissivity (e.g., Yakovlev et al. 2001). 
 
 The net effect of all this, for typical values of $S_v(n)$, is to give a $\beta$-equilibration 
 timescale $\tau^{\prime}_\beta \sim \tau_\beta/3$. From now on, we will use $\tau^{\prime}_\beta$ 
 and omit the prime.
 \subsubsection{Bulk viscosity regimes}
 \begin{figure*}
 \begin{center}
\includegraphics[scale=0.82]{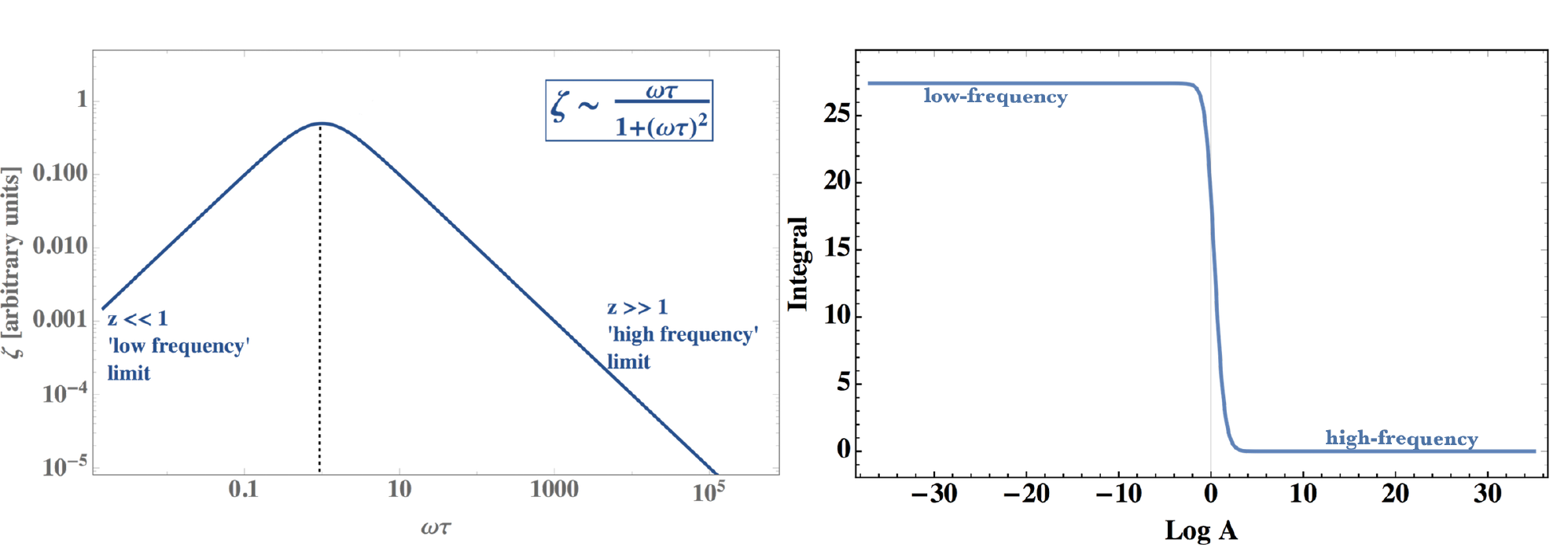}
\caption{{\it Left Panel:} Schematic representation of $\zeta(z)$ for
  a fixed perturbation frequency $\omega$;
  {\it Right Panel:} The $\xi$-integral of
  Eq.\ref{eq:energydiss-general} versus Log$A$, showing the
  change of $\dot{{\rm E}}^{({\rm bulk})}_{\rm diss}$ with bulk
  viscosity regime. The switch between regimes encompasses 3 orders of
  magnitude in $A$.}
\label{fig:sketch}
\end{center}
\end{figure*}
Defining the variable $z \equiv \omega \tau_\beta$, Eq. (\ref{eq:def-general}) implies two regimes of $\zeta$ as a function of $z$, 
as sketched in Fig. \ref{fig:sketch}  (left panel):
\begin{itemize}
\item[{\it i)}] $z \ll 1$, ``low frequency'' limit: $\beta$-reactions are much faster 
than the perturbation and chemical equilibrium is maintained almost instantaneously during one 
  oscillation. Deviations from equilibrium are thus tiny, and energy losses small. Accordingly, bulk viscosity is weak: 
  $\zeta \propto z$;
\item[{\it ii)}] $z \gg 1$, ``high frequency" limit: 
  $\beta$-reactions are much slower than the perturbation and, during one
  cycle, deviations from chemical equilibrium grow {\it almost}
  unimpeded. The effect of $\beta$-reactions builds up slowly,
  eventually dumping the perturbation over a large number of cycles: 
  $\zeta \propto z^{-1}$.
  \end{itemize}
 
\noindent
{\bf Pure $npe$ matter --}  In the 
high-frequency limit we have\footnote{Treating the NS as a collection
  of non-interacting, fully degenerate, fermion gases.}  (Sawyer 1989)
\be
\label{eq:bulkold}
\zeta_{\rm high} = 60~ T^6_{10} ~\frac{\rho^2}{\omega^2} \, .
\ee
The high-frequency limit holds as long as
\be 
\label{eq:high-frequency}
\cos \chi > \frac{P}{2 \pi \epsilon_B \tau_\beta} \approx
0.023~ T^6_{10}~\frac{{\rm P}_{\rm
    ms}}{\epsilon_{B, -3}} \left(\frac{\rho_{\rm
    n}}{\rho}\right)^{2/3} \, .  \ee For large ellipticities and
millisecond spins, {\it i.e.}, newborn magnetars, condition
(\ref{eq:high-frequency}) is always met unless the tilt angle is
$\approx \pi/2$. For smaller values of $\epsilon$ and longer spin
periods, expected for most NS, (\ref{eq:high-frequency}) is only
satisfied at sufficiently large angles and late times. Therefore,
ordinary NS start their life in the ``low-frequency" regime, switching
to high-frequency as they cool.

\medspace
\noindent
{\bf Realistic NS matter --} Adopting a more realistic EoS and chemical
composition, the bulk viscosity coefficient (\ref{eq:bulkold})  
can increase by a factor $N \sim 1.5-4$ (e.g. Haensel et al. 2001,
Dall'Osso \& Stella 2017). Accounting for this factor, and further
multiplying Eq.~(\ref{eq:bulkold}) by $z^2/(1+z^2)$, 
we derive a general expression for $\zeta$ as a
function of $z$, valid in any regime \be
\label{eq:general-zeta}
\zeta = \frac{60 N~ T^6_{10} ~\rho^2 z^2}{1+z^2} \approx
\frac{317 N \rho_{\rm n}^2 \left(\rho / \rho_{\rm
    n}\right)^{10/3} T^{-6}_{10}}{ \left[1+ \displaystyle \frac{5.3
      \epsilon_B^2 \Omega^2 \cos^2\chi \left(\rho/\rho_{\rm
        n}\right)^{4/3}}{T^{12}_{10}}\right]} \, .  \ee
In the following, the EoS-dependence of the bulk viscosity coefficient will be
simply parametrized by the value of $N$.

\subsubsection{Compressibility of fluid motions}
\label{sec:compressibility}
We turn now to the relation between $\Delta \rho$ and $\delta \rho$,
needed to calculate the energy dissipation rates 
(Eqs. \ref{eq:define-bulk-dissipation}, \ref{eq:define-shear-dissipation}).  Let us first recall that 
$\Delta \rho \equiv \delta \rho + \mathbf{\xi} \cdot
\nabla \rho$ and $\delta \rho \equiv -(\mathbf{\xi} \cdot \nabla \rho
+ \rho \nabla \cdot \mathbf{\xi})$, $\xi$ being the fluid displacement
due to the perturbation. When fluid motions are adiabatic, $\nabla
\cdot \mathbf{\xi} = \Delta \rho \equiv 0$: thus, one obtains $\delta
\rho = - \xi \cdot \nabla \rho$, which was used to calculate $\delta
\rho$ (Mestel \& Takhar 1972, Mestel et al. 1981).

It can be argued that, due to the periodically changing pressure
in the fluid, a field of motions with the magnitude calculated in the
adiabatic approximation will always be excited (e.g., Mestel \& Takhar
1972, Jones 1976, Lander \& Jones 2017). However, the compressibility of
such motions will change with the different physical regimes.  In the limit of 
highly dissipative fluid motions, for example, the density fluctuation $\delta \rho$ 
will occur mostly through a fluid compression, giving $\delta \rho \approx \rho \nabla
\cdot \mathbf{\xi}$ and, thus, $\mathbf{\xi} \cdot \nabla \rho \approx
0$. From this, we deduce $\Delta \rho \equiv \rho \nabla \cdot
\mathbf{\xi} \approx \delta \rho$.

\noindent
The relation between $\delta \rho$ and $\Delta \rho$ can thus be summarized as%depends on the regime of bulk viscosity. 
\begin{itemize}
\item[(i)] \textit{Low frequency, $\omega \tau_\beta \ll 1$}:  Because
  particle reactions are faster than the oscillation, density fluctuations are 
  accompanied by strong bulk compression of the fluid. The relation 
  $\Delta \rho \approx \delta \rho$ for highly dissipative motions can be
used in 
 % condition $\nabla \cdot \mathbf{\xi} \approx \Delta \rho/\rho$ will
 % hold. From this, the relation $\nabla \cdot \mathbf{\delta v}
  %\approx i \omega \Delta \rho/\rho$ is readily obtained, and
 (\ref{eq:define-bulk-dissipation}). This is the regime considered by Dall'Osso et al. (2009).

\item[(ii)] \textit{High frequency, $\omega \tau_\beta \gg 1$}: When
  particles reactions are slower than the perturbation, fluid motions are almost adiabatic. Thus, 
  $\Delta \rho < \delta \rho =  -(\mathbf{\xi} \cdot \nabla \rho)$, and the factor by which they differ will 
 be determined by the ratio between the two relevant timescales.
  %have $\Delta \rho  \ll \delta \rho$ 
 % we know that $\Delta \rho$ (hence
 % $\rho \nabla \cdot \mathbf{\delta v}$) should be smaller than that,
  %because $\Delta \rho \approx 0$ for almost adiabatic fluid flows. 
  In this regime we will adopt the relation (Dall'Osso \& Stella 2017) $\Delta \rho \approx \delta
  \rho (T_p/ \tau_\beta )$, where $T_p = 2 \pi / \omega$ is the
  precession period. Note that the ratio of timescales is a strongly decreasing function of time, since $T_p$ can only
  decrease (following the decrease of $\cos \chi$) while $\tau_\beta \propto T^{-6}$ is rapidly growing as the NS cools. Therefore,
  our expression describes the transition between the low-frequency and high-frequency regime and,  
  in the limit of a sufficiently low temperature, it tends to the condition $\Delta \rho \approx 0$ assumed by Lasky \& Glampedakis (2016).
  \end{itemize}
Following these arguments, we derive the general relation
\be
\label{eq:heaviside}
\Delta \rho \approx \delta \rho 
\left[\hat{\Theta}(T_p - \tau_\beta) + \frac{T_p}{\tau_\beta} ~\hat{\Theta}(\tau_\beta - T_p) \right] \equiv \delta \rho ~\hat{G}(T_p, \tau_\beta) \, ,
\ee
$\hat{\Theta}$ being the Heaviside function.
\subsubsection{Density perturbation}
\label{sec:density}
Mestel \& Takhar (1972) derived a general expression for the density
perturbation associated to the freebody precession of an oblique,
fluid rotator 
\be
\label{eq:deltarho}
\delta \rho \left(r, \Theta, \Phi, \Omega, \chi \right) = \frac{1}{2}
f(r) \hat{K}\left(\Theta, \Phi, \Omega, \chi \right)\,. 
\ee 
The angular
part $\hat{K}$, in which angles are defined {\it with respect to the
  magnetic pole}, is a complicated function to be discussed
later. The radial part is set uniquely by the NS EoS. Realistic NS EoS
can be approximated by piecewise polytropes with index $n \approx
0.5-1$, stiffening towards the center (Read et al. 2009). We verified
that the volume integral in Eq.~(\ref{eq:define-bulk-dissipation})
has a very weak dependence on $n$, slightly increasing for
stiffer EoS. For simplicity, we will assume $n=1$.
The adimensional density profile is $\hat{\theta} \left(\xi \right) =
\rho(\xi)/\rho_c$, where $\xi = r/\alpha$ is the radial
coordinate,  $\alpha = R_*/\pi$ and $\rho_{\rm c} = M/(4 \pi^2 \alpha^3)$ is the central density.

The density profile $\hat{\theta} (\xi)$ of a {\it rotating} polytrope
can be expressed, with respect to its non-rotating counterpart
$\hat{\theta}_0(\xi)$, in terms of the velocity parameter $v =
\Omega^2/(2 \pi \rho_c G)$ (Chandrasekhar 1933) \be
\label{eq:chandra-rot}
\hat{\theta}(\xi) = \hat{\theta}_0(\xi) + v\left[\psi_0(\xi) + A_2
  \psi_2(\xi) P_2(\cos \tilde{\Theta})\right] \, , \ee where $P_2(\cos
\tilde{\Theta})$ is a Legendre polynomial, and $\tilde{\Theta}$ the
latitude {\it with respect to the spin pole}.  The second term in
square brackets is the required non-spherical part of the rotational
perturbation. The function $\psi_2(\xi)$ can be calculated numerically
following Chandrasekhar (1933), and $A_2 \approx -0.54833$.

Recently, Lander \& Jones (2017) studied the same problem to a higher
perturbative order, including the effects on the magnetic field structure in the
fluid NS. The density perturbation that they derived is perfectly consistent with 
the one adopted here, both in amplitude, radial and angular dependence.

\subsubsection{The energy dissipation rate}
\label{sec:dissipation}
Inserting Eqs.~(\ref{eq:general-zeta}) and (\ref{eq:heaviside}) in
Eq.~(\ref{eq:define-bulk-dissipation}), we eventually derive 
the energy dissipation rate due to bulk
viscosity

\begin{eqnarray}
\label{eq:energydiss-general}
\dot{E}_{\rm diss} & = & \frac{317 N \alpha^3 A^2_2} {16 \pi^2 G^2}
\left(\frac{\rho_c}{\rho_{\rm n}}\right)^{4/3}
\frac{\Omega^6}{T^6_{10}} \epsilon^2_B \cos^2 \chi 
\nonumber \\ &\times &
\int_0^{\pi} d\xi
\frac{\xi^2 \psi^2_2(\xi)
  \left[\theta(\xi)\right]^{4/3} \hat{G}^2 (T_p, \tau_\beta)}{1+\displaystyle \frac{5.3
    \epsilon^2_B \Omega^2 \left(\rho_c / \rho_{\rm n}\right)^{4/3}
    \cos^2 \chi }{T^{12}_{10}} \left[\theta(\xi)\right]^{4/3}} 
\nonumber \\ & \times & \int d\Theta d\Phi~\hat{K}^2(\Theta, \Phi, \chi,
\Omega) \sin \Theta \, ,
\end{eqnarray}
where Eqs.~(\ref{eq:define-precession}, 
\ref{eq:deltarho}, \ref{eq:chandra-rot}) were used.  The angular integral, averaged over one precession period, 
is $24 \pi/5 \sin^2\chi (1+3 \cos^2\chi)$.

Expression (\ref{eq:energydiss-general}) follows the energy
dissipation rate as the NS switches from one regime of bulk viscosity
to the other. The number 1 in the denominator of
Eq. (\ref{eq:energydiss-general}) corresponds to the low-frequency
regime while the second term, which grows as the temperature drops
(although $\cos \chi$ and $\Omega$ decrease), represents dissipation
in the high-frequency limit. Note that
  Eq. (\ref{eq:energydiss-general}) depends on the NS EoS through
 $N$, as well as through $\alpha$ and $\rho_c$ (and hence mass and radius),
  appearing both in the normalization and inside the integral.

Writing the denominator in Eq.~(\ref{eq:energydiss-general}) as
$\left[1 + A \theta(\xi)^{4/3}\right]$, we calculated the integral
numerically for a wide range of values of Log$A$: results are shown
in the right panel of Fig.~\ref{fig:sketch}. %with the low-frequency regime 
%of bulk viscosity at small $A$ and the high-frequency regime at large $A$. 
%where the two regimes of
%bulk viscosity (low-frequency at small $A$, high-frequency at large$A$) are shown. 

For given NS parameters, and since $T(t)$ can be
calculated independently (Sec. \ref{sec:supercond}), our result gives
the integral in Eq.~(\ref{eq:energydiss-general}) as a function of
$\cos \chi$ and $\Omega$.  Finally, the damping timescale
  $\tau_{\rm d}$ is obtained by combining this expression for
  $\dot{E}_{\rm diss}$ with Eq.~(\ref{eq:precession-energy}), thus
  inheriting a dependence on M and R, as well as on the parameter $N$. These
  results will be used later (Sec. \ref{sec:evolution}) to solve
  numerically Eq.~(\ref{eq:evolution-equation}), for specific choices
  of the NS parameters.

\noindent
The asymptotes of Eq.~(\ref{eq:energydiss-general}) have the expressions
\begin{eqnarray}
\label{eq:asymptotics}
\dot{E}_1&=&\frac{951 \alpha^3 A^2_2} {10 \pi G^2} \left(\frac{\rho_c}{\rho_{\rm n}}\right)^{4/3} 
\frac{\Omega^6 \epsilon_B^2}{T^6_{10}} (\cos \chi \sin \chi)^2 (1+3 \cos^2 \chi)I_1  \nonumber \\
\dot{E}_2&=&\frac{317 \alpha^3 A^2_2} {10 \pi G^2} T^6_{10} \Omega^4  \sin^2 \chi (1+3 \cos^2 \chi) 
I_3 
\end{eqnarray}
where $I_1 \approx 17.6087 $, $I_2 \approx 148.815 $ for the $n=1$ polytrope.
\subsection{Shear viscosity}
\label{sec:shear}
The coefficient of shear viscosity in NS interiors was calculated in
detail by Shternin \& Yakovlev (2008). Unlike $\zeta$, the
coefficient $\eta$ grows as the temperature drops, and is further
increased by baryon condensation in the NS core. Therefore, since
shear viscosity dominates later stages of the NS life, we will
consider here only the expression for $\eta$ in a regime in which
protons are strongly superconducting while neutrons are still in a
normal state (see Sec. \ref{sec:supercond}) \be \eta \approx 10^{19}
\left(\frac{\rho_{15} x_{01}}{T_9^2} \right)^2 ~~~{\rm erg~cm}^{-1}\,
, \ee where $x_{01}$ is the proton fraction in units of 0.1.

The corresponding energy dissipation rate is obtained by integrating
Eq.~(\ref{eq:define-shear-dissipation}). To this aim, we recall the
discussion summarized in Eq.~(\ref{eq:heaviside}). The pressure/density
fluctuations produced by free body precession are achieved, in the
low-frequency limit ($z< 1$), mostly via fluid compression. In the
opposite regime, compression is very limited and density
fluctuations must be achieved by an almost adiabatic fluid circulation
(e.g., Mestel \& Takhar 1972).  The latter is the regime relevant
here: we can thus assume an almost adiabatic fluid circulation. In
this limit, the expression for $\delta \sigma$ in
Eq.~(\ref{eq:define-shear-dissipation}) must be, to order of magnitude,
$\sim \omega^2 \left(\delta \rho /\rho \right)^2$, as it was for
$\left| \nabla \cdot \delta {\mathbf{v}} \right|^2$ in the opposite
limit.  We will thus write \be
\label{eq:shear-dissipation}
\dot{{\rm E}}_{\rm diss}^{({\rm visc})} \approx 2 \times 10^{-11}
\frac{\omega^2}{T_9^2} \int x^2_{01} \delta \rho^2 dV \sim 10^{-7}
\frac{\omega^2}{ T_9^{23/3}} \dot{{\rm E}}_{\rm diss}^{({\rm bulk})}\,
.  \ee As a matter of fact, our substitution implies that the
shear-to-bulk viscosity dissipation timescale ratio is mostly
determined by the ratio $\zeta / \eta$, a conclusion already obtained
by, e.g., Cutler \& Lindblom (1987) in a different context. A general
conclusion from Eq.~(\ref{eq:shear-dissipation}) is that shear viscosity
can only affect the evolution of the tilt angle on timescales longer
than $10^7-10^8$ yrs, for temperatures $\gtrsim 2\times 10^8$ K,
ellipticities $\epsilon_B < 10^{-4}$ and spin periods $\gtrsim 10$
ms. Therefore, shear viscosity cannot affect the growth of $\chi$
before alignment driven by the electromagnetic torque kicks in.

\section{Superfluidity and superconductivity}
\label{sec:supercond}
The strong temperature dependence of viscous effects requires that we model
the NS cooling in order to calculate the long-term evolution of the
tilt angle. In particular, we must account for the transition to
superfluidity of baryons in the NS core. In addition to the effect
discussed in Sec. \ref{sec:shear}, superfluidity will reduce the rate of
$\beta$-reactions, implying (1) a decrease of the bulk viscosity
coefficient; (2) a decrease of the neutrino cooling rate, which will
keep the NS hotter than it would be otherwise. Neutron superfluidity can also affect the
precessional dynamics in significant ways (Shaham 1977, Sedrakian et al. 1999, Andersson et al.
2006). So, before proceeding further, we must specifiy the superfluid parameters that we assume, 
based on observational 
constraints derived from the cooling of the NS in Cas A
(Page et al. 2011, Shternin et al. 2011).
\begin{itemize}
\item[(i)] Neutron condensation (triplet state)
occurs at a critical temperature T$_{cn} \approx (5-6)
  \times 10^8$ K, which is reached at an age $\sim 300$ yrs in Cas
  A. Given that our calculations will extend up to $t \gtrsim $ 300 yrs, we
  will consistently 
  %Given that the effects of superfluidity are rather small as long as the temperature is $\gtrsim$ T$_{cn}/2$ (see below), we will 
   neglect neutron superfluidity: in particular, this implies that the neutrons making up most of the NS will precess as a ``normal" fluid. 
\item[(ii)] The proton energy gap (singlet state), $\Delta_{\rm p} \sim (0.5 - 1)$ MeV, implies a
  critical temperature T$_{cp} \sim (3.5-7) \times 10^9$ K. Thus,
  proton superconductivity occurs early in a NS life, and will be
  included in our model.
 \end{itemize}

\noindent
{\bf Reduction of bulk viscosity --} Haensel et al. (2001) provide analytical
 fits to numerical calculations of the reduction coefficient of bulk
 viscosity in NS matter due to baryon superfluidity.  We will consider their case with superconducting
 protons and normal neutrons and, for definiteness, we will set
 T$_{cp} = 5 \times 10^9$ K ($\Delta_p =0.75$
 MeV). Writing $\tau = T/T_{cp}$, the fitting formulae
 are\footnote{The subscript $p$ indicates that protons are
   superfluid.}
\begin{eqnarray}
\label{eq:reduction}
R^{(n)}_p &= &\frac{a^{5.5} +b^{3.5}}{2}~{\rm exp}\left[3.245-\sqrt{(3.245)^2+v^2}\right] \nonumber \\
R^{(p)}_p &=& c^5~{\rm exp}\left[5.033-\sqrt{(5.033)^2+(2v)^2}\right] \, ,
\end{eqnarray}
for the neutron ($n$) and proton ($p$) branch of the modified-Urca
reactions, respectively. The coefficients are
\begin{eqnarray}
\label{eq:coefficienti}
a & = & 0.1863+\sqrt{(0.8137)^2+(0.1310v)^2} \nonumber \\
b & = & 0.1863 + \sqrt{(0.8137)^2+(0.1437)^2} \nonumber \\
c & = & 0.3034 +\sqrt{(0.6966)^2+(0.1437v)^2} \nonumber \\
v & = & \sqrt{1-\tau} \left(1.456 -0.157/\sqrt{\tau}+1.764 /\tau\right) \, .
\end{eqnarray}

\noindent
{\bf Effect on NS cooling --}  Haensel et al. (2001) also provide analytical
fits to the reduction factor for the neutrino emissivity ($R_p^{\nu}$)
in superfluid NS cores.  This is slightly different from the coefficients in 
Eq.~(\ref{eq:reduction}). 

To model neutrino cooling of the NS we will consider three main
factors: (1) modified Urca reactions, the main emission process; (2)
proton superconductivity, which reduces the modified Urca reaction rate
by the factor $R_p^{(\nu)} <1$ at $T < T_{cp}$; (3) neutrino
bremsstrahlung, a weaker emission process that might become dominant
once proton superconductivity has suppressed modified Urca
reactions. \\

\noindent
Therefore, we will write \be
\label{eq:T-evol-with-typeII}
\frac{{\rm d}T}{{\rm d}t} = - \underbrace{\left[R^{(\nu)}_p(T)
    \frac{N^S}{C} + N^{({\rm br})}\right]}_{f(T)} T^7
\Rightarrow \int \frac{{\rm d}T}{f(T) ~ T^7} = -(t-t_0) \, ,  \ee
where $C_V(T) = C \cdot T$, and the coefficients on the r.h.s. are, in c.g.s. units, $N^S =10^{-32}$ and $C =10^{30}$ 
for modified Urca reactions, $N^{({\rm br})} =10^{-34}$ for neutrino
bremsstrahlung (e.g. Page et al. 2006).

\begin{figure}
\begin{center}
\includegraphics[scale=0.42]{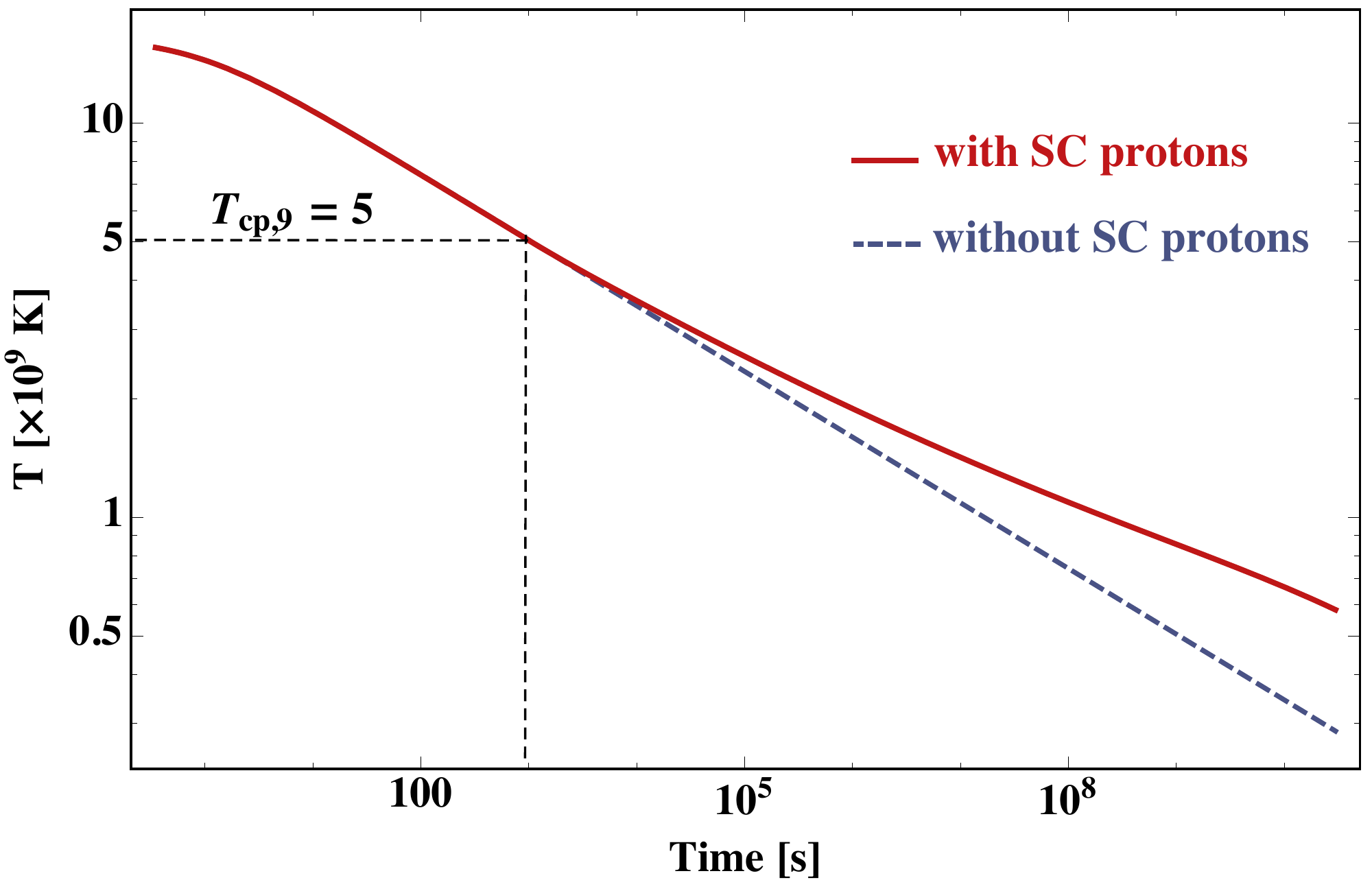}
\caption{Temperature evolution with and without proton
  superconductivity. The critical pairing temperature is T$_{cp} = 5
  \times 10^9$ K ($\Delta_p =0.75$ MeV).}
\label{fig:cooling}
\end{center}
\end{figure}

Fig. \ref{fig:cooling} shows the numerical solution of Eq. (\ref{eq:T-evol-with-typeII}), 
along with the cooling obtained from modified Urca reactions alone ($R_p^{(\nu)}=1$ and 
$N^{(\rm br)} = 0$). At $t \sim 10^{10}$ s, the age of Cas A, the red curve gives $T \sim 6 \times 10^8$ K, 
in good agreement with observational constraints. 

\section{Tilt angle distribution of newborn pulsars: effect of NS viscosity}
\label{sec:evolution}
{As already stated in Sec. \ref{sec:theoretical}, the
  pulsar population should be characterized by large tilt angles at
  birth, given the estimated alignment time $\sim 10^6-10^7$ yrs. With
  the classical dipole formula, the alignment timescale
  is\footnote{Note that $\tau_{\rm al}$ is a constant, since $\Omega
    \cos \chi$ is a conserved quantity.} (Jones 1976) \be
\label{eq:talign}
\tau_{\rm al} = \frac{2 \tau_{\rm sd,i}}{\cos^2 \chi_i}\;, 
\ee where
$\tau_{\rm sd,i} = \Omega_i/(2 \dot{\Omega}_i)$ is the NS spindown
timescale at birth, and $\chi_i$ the initial tilt angle. This relation implies that
the alignment time can be $\gg \tau_{\rm sd, i} \sim 10^3-10^5$ yrs
(for typical NS birth parameters), only if $ \chi_i \approx
90^\circ$. Recently, Philippov et al. (2014) have shown that (a) the
alignment time is somewhat longer when plasma effects in NS
magnetospheres are accounted for, obtaining $\tilde{\tau}_{\rm al} = 2
\tau_{\rm sd,i} \sin^2 \chi_i / \cos^4 \chi_i$, and (b) at late times,
$t \gg \tilde{\tau}_{\rm al}$, alignment further slows down, scaling
as $\sim t^{-k_2/2}$, where the structure constant $k_2 \approx
1$. Even in this case, a characteristic alignment time $\sim 10^6
-10^7$ yrs would require, at least, $\chi_i \gtrsim
60^\circ-70^\circ$.

\begin{figure*}
%\begin{center}
\includegraphics[scale=0.66]{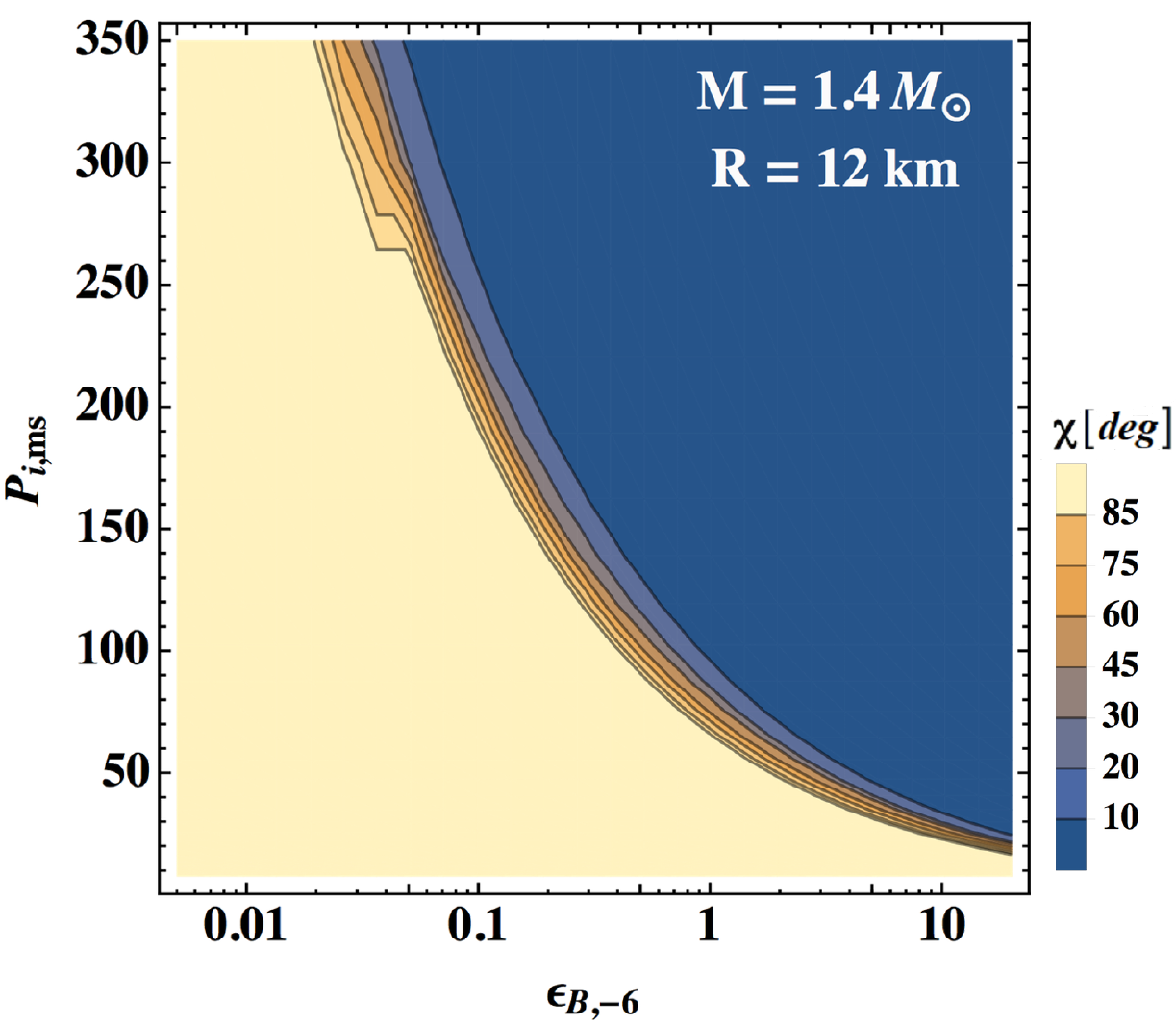}
\includegraphics[scale=0.66]{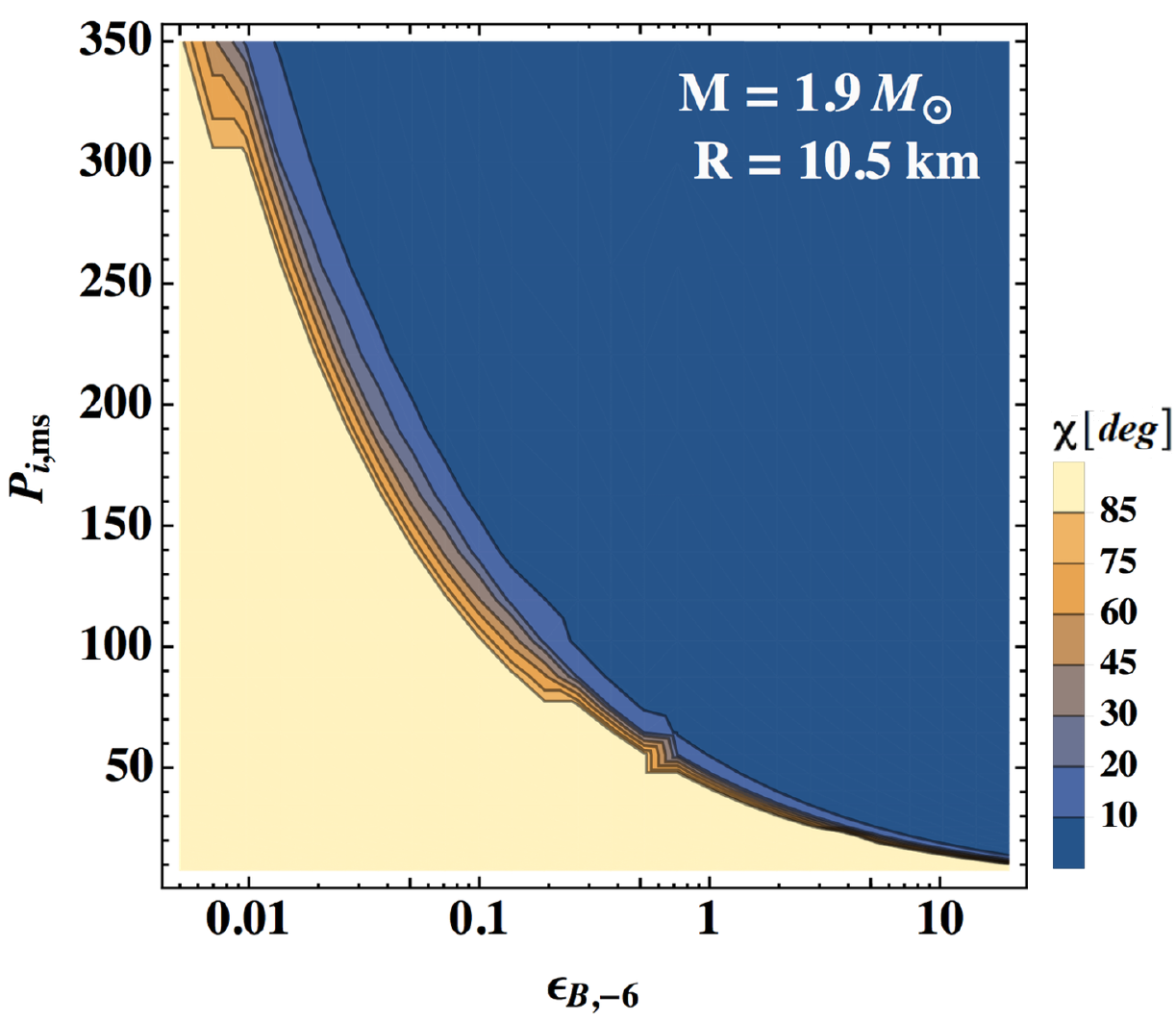}
\caption{The NS tilt angle $\chi$ at an age of  $10^{10}$ s, using
 M=$1.4 M_{\odot}$, R=12 km, N=3.5 (\textit{left} panel) or 
M=$1.9 M_{\odot}$, R=10.5 km, N=1.75 (\textit{right} panel), as a function of the initial 
spin period  $P_i$ and the magnetically-induced ellipticity $\epsilon_B$. The
  angle evolution occurs under the effect of the bulk and sheer
  viscosities, and it saturates after an age $ \lesssim $ a few years for all parameter choices 
  ($ \ll$ 1 yr in most cases). A remarkable result is the narrowness of the 
  parameter space $(P_i,  \epsilon_B)$ which leads to intermediate values of the tilt angle,
  $20^\circ\lesssim\chi\lesssim 70^\circ$. Growth saturation is
  achieved at either small or large angles, hence resulting in a
  bimodal distribution of tilt angles.}
\label{fig:contour}
%\end{center}
\end{figure*}

\noindent
It is therefore of great importance to be able to determine whether
tilt angles can grow enough in a timescale $< \tau_{\rm sd,i}$, since
this appears to be a general requirement of a long alignment time. If,
for example, the tilt angles were to remain relatively small in some
NSs, their alignment would be faster, potentially producing an
over-abundance of small tilt angles already at a young age
(cf. Rookyard et al. 2015a,b).

\subsection{Study of the parameter space}
\label{sec:parameter-space}
In order to explore the possible outcomes of the tilt angle evolution
on timescales shorter than $\tau_{\rm sd,i}$, we solve
Eq.~(\ref{eq:evolution-equation}) for a range of initial
conditions. We include the effects of bulk and shear viscosity and the
onset of proton superconductivity, and integrate the evolution
equations up to an age $t =10^{10}$ s.  Additional dissipative
processes, which might affect the growth of $\chi$ on longer times (see
Sec. 6 and 7), will be considered in future work, along with the
long-term alignment driven by the electromagnetic torque.  

Results of the evolution turn out to be quite sensitive to the 
choice of the NS EoS, and hence on mass, radius and $N$. Therefore, we chose two
cases that encompass the range of uncertainty in NS parameters: 1) a relatively
low-mass, large-radius NS (1.4 M$_{\odot}$, 12 km) with a relatively large value of the bulk viscosity 
coefficient ($N =3.5$); 2) a relatively high-mass NS (1.9 M$_{\odot}$, 10.5 km), with 
a smaller value of the bulk viscosity coefficient ($N=1.75$).

Note that, since $\tau_{\rm d}$ depends on the NS spin for either type of
viscosity considered here, Eq. ~(\ref{eq:evolution-equation}) is
coupled to the spin evolution of the NS. The latter is determined by
the dipole formula (Spitkovsky 2006): \be
\label{eq:spit}
 L_{\rm sd} = \mu^2/c^3 \Omega^3 \left(1+ \sin^2 \chi \right) \, , 
\ee
 where $\mu = B_p/2 R^3$ is the magnetic dipole moment and $B_p$ the
 dipole field strength at the magnetic pole.}

\noindent
Once the microphysics is specified (Sec. \ref{sec:viscosity-NS} and
\ref{sec:supercond}), the relevant parameters of the model are the NS
birth spin period, $P_i$, and the magnetically-induced ellipticity,
$\epsilon_B$. {We cover a wide range of $P_i$ and
  $\epsilon_B$, inclusive of plausible NS initial spin and ellipticity values that would make the
  magnetic deformation dominant over other possible sources. The
  magnetic dipole is fixed at a typical value B $\sim 3 \times
  10^{12}$~G, since NS spindown occurs on timescales longer than the
  viscous effects considered here. This is true as long as B~$ <
  10^{13}$~G and $P_i > 10$ ms: these values set the limit of validity
  of our study.  Cases not included here, with stronger B and/or
  faster spin, have a more rapid spindown, which slows down viscous
  dissipation (cf. Eq. \ref{eq:energydiss-general}) and accelerates
  the alignment due to the electromagnetic torque. Both effects make
  it more likely that the tilt angle of a fast spinning, highly
  magnetized NS remains small: it can grow to large values, though, if the 
  B-field in the NS core is particularly strong (cf. Dall'Osso et al. 2009, Dall'Osso \& Stella
  2017).}

Fig. \ref{fig:contour} shows the value of the tilt angle, as a
function of $P_i$ and $\epsilon_B$, at two different ages.  In the
left panel we consider a very young age, $\sim 10^4$ s, where little
evolution of the tilt angle occurs apart from a small region in
parameter space. In the right panel the tilt angle is shown at a much
later age, $\sim 300$ yrs, and we can appreciate the dominant effect
of viscous evolution. Note that the effect of the
  magnetic dipole on the tilt angle is still negligible at this
  age. The effect of viscosity, on the other hand, is manifested much
  earlier: even for the slowest-evolving NS in our grid, the tilt
  angle stops growing at $t \lesssim $ 10 yrs, since bulk viscosity is
  quenched after that time.

It is important to notice that there is only a small fraction of the
parameter space where the tilt angle has intermediate values. In
general, viscous evolution {\it appears to prefer either small angles
  or full orthogonalization}. This bimodality is mainly a consequence
of the strong temperature-dependence of the bulk viscosity, which
kills off viscous evolution once the NS has cooled below $(1-2) \sim
10^9$ K. After that point, tilt angles in our model do
  not change substantially, as the effect of shear viscosity is
  negligible on the timescales of interest. They might of course be
  affected by additional sources of viscosity, not included in our
  current model and working on longer timescales: one such
possibility will be discussed in Sec. \ref{sec:Crab}. However, we
leave a systematic study of these additional effects to future work.

\subsection{Pulsar observations}
\label{sec:observations}
Compilations of pulsar tilt angles (Tauris \& Manchester 1998, Young et al. 2010, 
Rookyard et al. 2015a) show an indication for a bimodal distribution of tilt angles, with 
two peaks at $\lesssim 40^{\circ}$ and $\gtrsim 80^{\circ}$. As these authors discuss, 
observational biases favour the detection of highly inclined objects: thus, the abundance
of pulsars with tilt angles $ \leq 40^\circ$ is particularly relevant,
as is the lack of objects with intermediate angles ($\sim
40^\circ-80^\circ$), given that orthogonal rotators ($\alpha >
80^\circ$) are fairly numerous. 

A comparison with the tilt angles of magnetars is less direct, since
magnetars have been seen to emit in radio only following an outburst
(Camilo et al. 2006), and the location of the outburst may not be the
same as that of the dipolar field. Using quiescent X-ray data,
constraints have been made on the viewing angle (i.e. the angle between 
the line of sight and the hottest region on the star) in several objects
(i.e. Dedeo et al 2000; Perna \& Gotthelf 2008; Bernardini et
al. 2011; Guillot et al. 2015).  However, if toroidal fields are
largely dominant in the NS crust (Thompson \& Duncan 1995; Perna \&
Pons 2011), then the location of the hottest point on the surface of
the star may not be coincident with that of the magnetic pole (Vigan\'o
et al. 2013; Perna et al 2013), and hence inferences of the viewing
geometry may not yield the correct value of the tilt angle.
Therefore, here we only consider the observational data set of the `standard'
pulsars, whose observations in radio and gamma-rays allows a more
reliable inference of the distribution of the tilt angles. 

However, even with this sample we emphasize that, while the observed
angle distributions are highly suggestive that our proposed mechanism
might be important in pulsars, a more quantitative comparison with
observations is not possible at this stage. First, observational
biases should be accounted for to translate our calculated
distribution into an observed one. Second, NSs in different samples
have ages in the range $10^3-10^8$ yr: as already stated, additional
sources of viscosity and the alignment caused by magnetic dipole  
should be accounted for explicitly on such long timescales.  Nevertheless, the results of
Fig. \ref{fig:contour} appear very encouraging for our scenario. In
the next section, taking it a step further, we will apply our
formalism to the special case of the Crab pulsar, where measurements
allow to test the nature of additional viscous processes.

\subsection{Corollary: oblate ellipsoids}
\label{sec:oblate}

We briefly comment on a crucial assumption made in
Sec. \ref{sec:theoretical}. Viscous damping of free precession will
work even in the absence of a strong toroidal B-field in the NS
core. In this case, the dipole B-field would cause an {\it oblate}
distortion, which would be dominant as long as $B_p > 3 \times 10^{11}
\left({\rm P}_{\rm ms}/10\right)^{-2}$ G, for superconducting protons
(Cutler 2002). In an oblate ellipsoid, minimization of the rotational
energy at constant angular momentum will align the NS symmetry axis with the
spin axis and, as a result, will cause a decrease of the tilt angle
towards zero.  Alignment will still occur on the viscous timescale,
according to Eq. \ref{eq:evolution-equation}. For the typical magnetic
field of pulsars, the ellipticity would be $\sim 10^{-10}-10^{-9}$,
implying a particularly short timescale for damping through bulk
viscosity. We solved Eq. \ref{eq:evolution-equation} for a few values
of the initial tilt angle and birth spin, and verified that NSs with a
predominantly oblate shape distortion would become aligned rotators
very early in their lives, unless they were born with tilt angles
$\approx 90^\circ$. Thus, dismissing the idea of a prolate distortion
poses an even greater problem, given that pulsars are not all
aligned. Alignment would require that the SN mechanism produces
preferentially orthogonal rotators.

An oblate shape, and viscous alignment, would also result if the NS
ellipticity was determined by the elastic deformation of the crust. In
this case, precession would involve the NS crust, while crust-core
coupling would produce dissipation, likely on longer
timescales. However, the elastic deformation can exceed the magnetic
one only for very low magnetic fields (see above): this case is
unlikely to be relevant for the bulk of the NS population.

\section{Crab Pulsar}
\label{sec:Crab}
The Crab pulsar has a spin period P$_C\simeq$ 33.7 ms and period
derivative $\dot{P}_C \simeq 4.23 \times 10^{-13}$ (Abbott et al. 2008). With the dipole
formula (\ref{eq:spit}) and $\chi_C \approx 60^\circ$ (Harding et
al. 2008; Watters et al. 2009; Du, Qiao \& Wang 2012), we estimate
B$_{\rm p} \simeq (3.3-5.2) \times 10^{12}$ G for the mass/radius range of Sec. \ref{sec:evolution}. 
The tilt angle is observed to be growing at the rate $\dot{\chi}_{\rm C} \approx 0.62^\circ$/century
or $3.6 \times 10^{-12}$ rad s$^{-1}$ (Lyne et al. 2013).

The measured growth rate for the tilt angle has been interpreted in
terms of freebody precession of the NS, which would require its
symmetry axis to be almost aligned with either the spin or the
magnetic axis (Philippov et al. 2014, Zanazzi \& Lai 2015), since
$\dot{\chi} \ll \Omega \epsilon$ even for the smallest possible
deformation of the NS. In this interpretation, the growth of $\chi$
is only an ``apparent" effect, associated to one half of the
precession cycle (period $\sim 100-200$ yrs): the secular trend can
only be an alignment, driven by the electromagnetic torque.
\subsection{Viscous damping of precession} 
\label{sec:damping}
Here we consider the alternative, that $\dot{\chi}_C$ effectively
represents a growth of the tilt angle, driven by a slow dissipative
process in the NS core. This of course requires that the NS has a
predominantly prolate shape, hence a toroidal magnetic field in its
core. From Fig. \ref{fig:contour}, and for the likely birth spin\footnote{Obtained from self-consistent solutions to 
Eqs.~\ref{eq:evolution-equation} and \ref{eq:spit}.} $\gtrsim $ 20~ms (e.g. Haensel et al. 2007), we see that 
$\epsilon_B \sim (3-10) \times 10^{-6}$ in order for the tilt angle
  %\footnote{Note that this specific range of values is strictly related to the fact 
  %that the tilt angle is well below $ 90^\circ$. $\epsilon_B$ could
%        be smaller, or even much smaller.}}}} 
to be in the narrow range $\sim (45^\circ - 75^\circ)$. Note that the low-end of the range corresponds to the
more massive NS, having a smaller radius and a lower $N$-value, while
the high-end is reached in the opposite case. In the superconducting
NS core, Maxwell stresses are enhanced by a factor $H_{c1}/B$, where
the critical field $H_{c1} \sim 10^{15}$ G.  This yields the
scaling $\epsilon_B \sim 5 \times 10^{-9} H_{c1,15} B_{12}$
(e.g. Cutler 2002), from which the (volume-averaged) toroidal magnetic
field is estimated, B$_T \sim (6-20) \times 10^{14}$ G, for M$= 1.4$
M$_{\odot}$ and R = 12 Km.
\noindent
In this intepretation, the measured $\dot{\chi}_C$ implies, via
Eq.~(\ref{eq:evolution-equation}), a dissipation time $\tau_{\rm d} \sim
3 \times 10^{11} \cos \chi/\sin \chi \approx (0.5 -3) \times 10^{11}$~s,
too short for either bulk or shear viscosity. This is also too
fast, and of the wrong sign, to be consistent with the effects of the
electromagnetic torque.

An additional viscous process consistent with the above estimate could
be crust-core coupling via mutual friction in the superfluid
core\footnote{At the age of the Crab, both protons and neutrons are
  expected to be in a condensed state.}, if the NS has a triaxial
shape.  The mutual friction coupling time is estimated to be (e.g.,
Jones 2012 and references therein) \be
\label{eq:mutual}
\tau_{\rm mf} = \frac{1}{{\cal R} \Omega \epsilon \cos \chi} \frac{I_{\rm
    pre}}{I_{\rm sf}} \approx 5 \times 10^{10} \frac{{\rm P}_{\rm
    ms}/33}{\epsilon_{-10}~\cos \chi}~\frac{I_{\rm pre}/I_{\rm
    sf}}{0.03} ~{\rm ~s} \, , 
\ee 
where we have used ${\cal R} \approx
5 \times 10^{-5}$ (e.g. Ashton et al. 2017) and $I_{\rm pre}/I_{\rm sf}$ is
the crust-core ratio of moments of inertia.  An ellipticity $\epsilon
\sim 10^{-10}-10^{-9}$ matches well the number expected from the
elastic deformation of the NS crust, if either set by the centrifugal
force 
\be 
\label{eq:centrifuge} 
\epsilon^{{\rm (cfg)}}_{\rm el} \sim b \epsilon_0 \approx 10^{-7}
\epsilon_0 \approx 5 \times 10^{-11} \left(\frac{{\rm P}_{\rm
    ms}}{33}\right)^{-2} \, , \ee or by the crustal breaking strain
(e.g., Jones 2012) \be \epsilon^{{\rm (break)}}_{\rm el} \sim b
u_{break} \approx 10^{-9} \left(\frac{u_{break}}{10^{-2}}\right)\, .
\ee How does this make the tilt angle grow, given that the elastic
deformation produces an oblate ellipsoid?  Let us first consider the
fluid interior: here, the centrifugal deformation is always aligned
with the $\Omega$-axis, while the magnetic deformation is tilted by
the angle $\chi$: it is the latter that excites freebody
precession. The NS crust inherits the same magnetically-induced
deformation, from the time when it first crystallized. Thus, if there
was only the magnetic deformation, the NS core and crust would precess
together, at the frequency $\omega = \epsilon_B \Omega \cos \chi$, and
no friction would occur. However, because of the magnetic tilt and of
crustal elasticity, a small part of the centrifugal deformation in the
crust is misaligned with respect to the $\Omega$-axis: this gives rise
to $\epsilon_{\rm el}$, an extra ellipticity specific to the crust.
Therefore, in the frame of the NS core, the crust will have an extra
``periodic" motion associated to $\epsilon_{\rm el}$\footnote{The
  motion of the triaxial crust is, in general, more complicated. 
  Our argument should be considered valid as an order of magnitude estimate.}: it is this extra
motion that induces mutual friction, and dissipation on the timescale
(\ref{eq:mutual}). If, however, the NS has a {\it prolate} magnetic
ellipticity $\epsilon _B \gg \epsilon_{\rm el}$, then the minimum
energy state will have the magnetic axis orthogonal to the spin axis,
a condition that will also guarantee alignment of the centrifugal
deformation in the crust with the $\Omega$-axis. It is interesting to
mention here the suggestion (Jones et al. 2016) that a triaxial star,
with both crustal and magnetic deformation (and, possibly, a prolate
shape), might help resolve a tension between glitches and precession
in PSR B1828-11.

Two comments are in order concerning the constrain on $\epsilon_B$ implied by our model. First, the required toroidal magnetic field in the Crab
is two orders of magnitude stronger than the poloidal one. As already stated (Sec. \ref{sec:theoretical}), the question of magnetic equilibria in NS interiors is an open issue in 
current research, with results pointing towards possible poloidal-dominated states (e.g., Colaiuda et al. 2008, Lander \& Jones 2009, Lander 2013) or toroidal-dominated ones (Spruit \& Braithwaite 2004, Braithwaite 2009, Pons \& Perna 2011, Akg\"{u}n et al. 2013). Our result is consistent with the stability limit of toroidal fields derived for a stably stratified NS interior (Agk\"{u}n et al. 2013, Dall'Osso et al. 2015). From this point of view, the tilt angle of the Crab pulsar and its measured growth rate might represent a signature of specific physical conditions existing in the NS interior, not accessible to direct observation.

Second, we note that the constraint on $\epsilon_B$ found above is a factor $\sim$ 3-10 lower than the most recent upper limits placed by 
the first science run of Advanced LIGO (Abbott et al. 2017).  This makes it possible that the Crab pulsar may become an interesting 
source of GWs for current laser interferometers, once they operate at design sensitivity. The orientation-averaged, 
instantaneous strain for a NS spinning at frequency $\nu$, with an ellipticity $\epsilon$ and at a distance $d$ is (Ushomirsky, 
Cutler \& Bildsten 2000)
\begin{eqnarray}
h_a  & = &4 \pi^2 \sqrt{\frac{2}{5}}  \frac{G}{c^4} \frac{I \epsilon}{d} \nu^2 \sqrt{\sin^2 \chi (1+15 \sin^2 \chi)} \nonumber \\
 &\approx & 1.4 \times 10^{-27} \frac{\epsilon_{-6}}{(d/2 {\rm kpc})} \left(\frac{{\rm P}_{\rm ms}}{33}\right)^{-2}
\end{eqnarray}
for a tilte angle $\chi = \pi/3$, where the GW signal frequency is $f = 2 \nu$. For a total observing time $T$, 
%the characteristic strain can be estimated as $h_c = h_a \sqrt{N}$, where$N = T f$ is the total number of cycles that have been observed, and 
the minimum detectable amplitude for a single template search with ${\cal D}$ detectors at frequency $f$ is $h_0^{({\rm min})} = 11.4 \sqrt{S_h(f)/{\cal D} T}$ (Andersson et al. 2011), where $S_h$ is the one-sided noise spectral density of the detector at frequency $f$. Therefore, the required observing time $T$ in order for the GW signal to be detectable
can be estimated roughly by setting $h_a > h_0^{({\rm min})}$. Plugging
in the numbers for the Crab, and adopting the design sensitivity curve of Advanced LIGO at\footnote{It is $S_h(60 Hz) \approx 2.5 \times 10^{-47}~{\rm Hz}^{-1}$ (e.g. Martynov et al. 2016).} $f \approx 60$ Hz, we obtain
\be 
T > \frac{6 \times 10^7~{\rm s}}{{\cal D}}~ \frac{({\rm P}_{\rm ms}/33)^{4} (d/ 2~{\rm kpc})^2}{(\epsilon_{-6}/5)^{2}}  \, .
\ee
%corresponding to 5-6 months worth of data. 

\subsection{Further implications}
\label{sec:implications}
The measured value of $\dot{\chi}_C$ suggests that additional sources
of viscosity, in addition to those considered in
Sec.~\ref{sec:viscosity-NS}-\ref{sec:evolution}, may affect the tilt
angle on timescales $> 10^3$ yrs.
Interpreting $\dot{\chi}_C$ as due to mutual friction, the expected
tilt angle evolution can be calculated by inserting (\ref{eq:mutual})
and (\ref{eq:centrifuge}) into (\ref{eq:evolution-equation}), which
gives $\dot{\chi} \propto \Omega^3 \cos^2 \chi / \sin \chi$. Thus, the
growth rate of the tilt angle decreases rapidly with time, as the NS
spins down. Because mutual friction is due to the interaction, in the
NS core, between the charged particles\footnote{Coupled to the crust
  on a very short timescale.} and the superfluid neutron vortices,
expression (\ref{eq:mutual}) will only hold at $T < T_{cn}$, {\i.e.}
at an age $> 300$ yrs if the cooling of the NS in Cas A can be taken
as representative. Hence, starting from $t_i= 10^{10}$ s, one may
estimate the overall effect of mutual friction in the Crab pulsar by
the integral $\Delta \chi = \int_{t_i}^{t_F} \dot{\chi}
dt$. Normalizing to the current value $\dot{\chi}_C$, and
approximating\footnote{This overestimates the integral given that
  $\chi$ actually grows.}  $\cos^2 \chi/\sin \chi \approx \cos^2
\chi_C/\sin \chi_C \approx 0.3$, we obtain $\Delta \chi \lesssim
19^\circ$, for $t_F \gg \tau_{\rm sd,i}$ and an initial spin period
$\sim 20$ ms. Thus, mutual friction would have a limited, yet non
negligible, effect on the long term evolution of the tilt angle.

The above estimate is an absolute upper limit to $\Delta
\chi$. Indeed, for $\chi_C \sim 60^{\circ}$, the alignment time due to the 
electromagnetic torque is $\tilde{\tau}_{\rm al} \sim 24 \tau_{\rm sd,i} \sim 2 \times 10^4$ yrs 
 (Sec. \ref{sec:evolution}). Thus, this mechanism would give an
important {\it negative} contribution to $\dot{\chi}$, limiting the
growth of $\chi$ already at an early age before causing its decrease
 on longer timescales.

We can draw a general lesson from the above argument. Slow dissipative processes, like e.g. mutual friction, can
offset the tilt angles calculated in Sec.~\ref{sec:evolution} by non-negligible amounts. However, they are unlikely
to substantially alter the bi-modality in the tilt angle distribution shown in Fig.~\ref{fig:contour}. 
Such slower processes, and the alignment due to the electromagnetic torque, depend on additional
physical parameters that cannot  be fully modelled at this stage and hence
will have to be reserved to future investigations.
 
\section{Conclusions}
\label{sec:conclusions}
We have studied the effect of viscosity on the tilt angle evolution of
newborn NSs. In particular, we have modelled bulk and shear viscosity
of NS matter, following the neutrino cooling down to a temperature T
$\sim 10^9$ K. Effects of proton superconductivity on both types of
viscosity and on the NS cooling are accounted for, assuming $ T_{cp}
\sim 5 \times 10^9$ K.  We have focused on a specific scenario in
which (i) the NS has a non-spherical shape which is mostly determined
by magnetic stresses, (ii) the internal magnetic field is dominated by
a toroidal component, which causes a prolate deformation, and (iii) at
birth, the magnetic axis has a small tilt angle (with respect to the
spin axis), which grows by viscous dissipation of free precession, the
latter being excited by the magnetic distortion.  We have solved the
evolution equation for the tilt angle (\ref{eq:evolution-equation}),
coupled to the spin evolution of the NS, for a wide range of values of
the initial spin ($P_i \sim 10-300$~ms) and the magnetically-induced
ellipticity ($\epsilon_B \sim 10^{-8} - 10^{-5}$), up to an age of
$10^{10}$~s. At this age, all viscous effects are already exhausted,
while the magnetic dipole-driven alignment of the magnetic axis has
not yet started and neutron superfluidity has not yet occurred (for
$T_{cn} \lesssim 6 \times 10^8$ K).

Our results show that viscous evolution of the tilt angle can either
lead to fast orthogonalization of the NS magnetic axis, or to very
little evolution, depending on the combination of $P_i$ and
$\epsilon_B$. For most parameter combinations, the tilt angle at the
end of the integration is either $\gtrsim 80^\circ$ or $ \lesssim 10^\circ$:
the slower the NS spin, the larger the ellipticity required for
$\chi$ to reach $\approx 90^\circ$. Parameter combinations that lead
to intermediate values of the tilt angle occupy a relatively narrow
strip in parameter space. This very pronounced bi-modality is a result
of bulk viscosity dominating the evolution: NSs start their life in the
low-frequency regime, where the bulk viscosity coefficient is very
low, and later evolve towards the high-frequency regime as they cool. Most
of the dissipation occurs around the turnover, where the bulk
viscosity coefficient has a peak, which roughly sets the timescale for
orthogonalization via the condition $\omega \tau \sim 1$. The latter
is essentially a relation between $\epsilon_B$ and $P_i$, with
temperature (time) as a parameter.

Our model does not include additional viscous processes, that might
affect the tilt angle evolution on longer timescales. For example,
crust-core coupling is an important aspect that we have not addressed
here. Processes of this type could help populate the region of
intermediate tilt angles, on timescales $ > 10^{10}$ s but still short
compared to the alignment time. As such, they may mitigate the
pronounced bi-modality in tilt angle distribution expected from our
results (Fig. \ref{fig:contour}), yet without removing it. This
possibility, and its implications for the long-term alignment, will
have to be addressed in future studies.

We have applied our model, in particular, to the Crab pulsar, where
measurements of $\chi$ and $\dot{\chi}$ allow a more direct test of
theoretical expectations.  If due to viscous dissipation, the measured
growth of the tilt angle implies that the NS has a predominantly
prolate deformation, hence a toroidal magnetic field. The fact that
the tilt angle is less than $90^\circ$, and still growing, implies
that the magnetically-induced ellipticity should be $\gtrsim 3 \times
10^{-6}$ for a likely birth spin $\sim20$~ms
(Fig. \ref{fig:contour}). The corresponding toroidal magnetic field is
$\gtrsim 6 \times 10^{14}$ G (assuming protons are superconducting),
about 100 times larger than the large-scale dipole. The measured vale
of $\dot{\chi}$, on the other hand, points to a dissipation time $\sim
(0.5 - 3) \times 10^{11}$ s, the physical interpretation of which is
still open. With bulk and shear viscosities ruled out, crust-core
coupling via mutual friction would be a natural candidate: the
relative motion between core and crust might be due to the elastic
deformation of the latter. In this case, mutual friction would work on
a timescale set by $\epsilon_{\rm el} \sim \times 10^{-10} -10^{-9}$,
but the growth of $\chi$ is guaranteed by the larger,
magnetically-induced distortion. 

This scenario has two possible observational tests: one is the
existence of modulations in the timing residuals of the Crab pulsar, with
a period of $\sim (0.4-2) \times 10^5$ s or $\sim 3 \times 10^8$ s,
associated to either $\epsilon_B$ or $\epsilon_{el}$. Detection of the
faster modulation might not be straightforward, though, if the radio
beam is approximately aligned with the magnetic axis and, hence, with
the precession axis. Given the geometrical constraints, the slower
modulation might be more easily detectable. The second, even more
direct, test is the detection of a periodic GW signal at $\nu \approx
60$~Hz, due to the magnetic deformation $\epsilon_B \gtrsim
10^{-6}$. In this case, the instantaneous strain would be $h \gtrsim 10^{-27}$ for a distance of 2 kpc, 
%corresponding to a characteristic strain (e.g., Moore et al. 2015) $ h_c \gtrsim 10^{-23}$ for a $\sim$ 70 day worth of data (as in 
%the Advanced LIGO O1, Abbot et al. 2017), 
likely detectable by Advanced LIGO/Virgo once operating at design sensitivity.  ~\\ 

{\bf Acknowledgments}.
This work was supported by NSF award AST-1616157.


\begin{thebibliography}{}
\bibitem[Abbott et al.(2008)]{} Abbott, B., Abbott, R., Adhikari, R., et al.\ 2008, \apjl, 683, L45
\bibitem[Abbott et al.(2017)]{} Abbott, B.~P., Abbott, R., Abbott, T.~D., et al.\ 2017, \apj, 839, 12
\bibitem[Akg{\"u}n et al.(2013)]{} Akg{\"u}n, T., Reisenegger, A., Mastrano, A., \& Marchant, P.\ 2013, \mnras, 433, 2445
\bibitem[Alpar \& Pines(1985)]{} Alpar, M.~A., \& Pines, D.\ 1985, \nat, 314, 334
\bibitem[Alpar \& Sauls(1988)]{} Alpar, M.~A., \& Sauls, J.~A.\ 1988, \apj, 327, 723 
\bibitem[Andersson et al.(2006)]{} Andersson, N., Sidery, T., \& Comer, G.~L.\ 2006, \mnras, 368, 162
\bibitem[Andersson et al.(2011)]{} Andersson, N., Ferrari, V., Jones, D.~I., et al.\ 2011, General Relativity and Gravitation, 43, 409
\bibitem[Ashton et al.(2016)]{} Ashton, G., Jones, D.~I., \& Prix, R.\ 2016, \mnras, 458, 881
\bibitem[Ashton et al.(2017)]{} Ashton, G., Jones, D.~I., \& Prix, R.\ 2017, \mnras, 467, 164
\bibitem[Baym et al.(1969)]{} Baym, G., Pethick, C., \& Pines, D.\ 1969, \nat, 224, 673
\bibitem[Bernardini et al. (2011)]{} Bernardini, F., Perna, R., Gotthelf, E. Israel, G. L. Rea, N., 
Stella, L. 2011, \mnras, 418, 638
\bibitem[Braithwaite \& Spruit(2004)]{} Braithwaite, J., \& Spruit, H.~C.\ 2004, \nat, 431, 819
\bibitem[Braithwaite(2006)]{} Braithwaite, J.\ 2006, \aap, 449, 451
\bibitem[Braithwaite \& Nordlund(2006)]{} Braithwaite, J., \& Nordlund, {\AA}.\ 2006, \aap, 450, 1077 
\bibitem[Braithwaite(2009)]{} Braithwaite, J.\ 2009, \mnras, 397, 763 
\bibitem[Chandrasekhar(1933)]{} Chandrasekhar, S.\ 1933, \mnras, 93, 390 
\bibitem[Colaiuda et al.(2008)]{} Colaiuda, A., Ferrari, V., Gualtieri, L., \& Pons, J.~A.\ 2008, \mnras, 385, 2080
\bibitem[Cutler \& Lindblom(1987)]{} Cutler, C., \& Lindblom, L.\ 1987, \apj, 314, 234
\bibitem[Cutler(2002)]{} Cutler, C.\ 2002, \prd, 66, 084025
\bibitem[Dall'Osso et al.(2009)]{} Dall'Osso, S., Shore, S.~N., \& Stella, L.\ 2009, \mnras, 398, 1869 
\bibitem[Dall'Osso et al.(2012)]{} Dall'Osso, S., Granot, J., \& Piran, T.\ 2012, \mnras, 422, 2878
\bibitem[Dall'Osso et al.(2015)]{} Dall'Osso, S., Giacomazzo, B., Perna, R., \& Stella, L.\ 2015, \apj, 798, 25 
\bibitem[Dall'Osso \& Stella (2017))]{} Dall'Osso, S., Stella, L., 2017 {\it in preparation}
\bibitem[DeDeo et al. (2000)] {} DeDeo S., Psaltis D., Narayan R., 2001, \apj, 559, 346
\bibitem[Du et al.(2012)]{} Du, Y.~J., Qiao, G.~J., \& Wang, W.\ 2012, \apj, 748, 84
\bibitem[Easson \& Pethick(1977)]{} Easson, I., \& Pethick, C.~J.\ 1977, \prd, 16, 275 
\bibitem[Friedman \& Stergioulas(2013)]{} Friedman, J.~L., \& Stergioulas, N.\ 2013, Rotating Relativistic Stars, by John L.~Friedman , Nikolaos Stergioulas, Cambridge, UK: Cambridge University Press, 2013
\bibitem[Goldreich(1970)]{} Goldreich, P.\ 1970, \apjl, 160, L11
\bibitem[Guillot et al. (2015)]{} Guillot, S., Perna, R., Rea, N., Vigan\'o, D., Pons, J. A.
2015, \mnras, 452, 3357 
\bibitem[Haensel et al.(2001)]{} Haensel, P., Levenfish, K.P., Yakovlev, D.G.\ 2001, \aap, 372, 130
\bibitem[Haensel et al.(2007)]{} Haensel, P., Potekhin, A.~Y., \& Yakovlev, D.~G.\ 2007, Astrophysics and Space Science Library, 326
\bibitem[Harding et al.(2008)]{} Harding, A.~K., Stern, J.~V., Dyks, J., \& Frackowiak, M.\ 2008, \apj, 680, 1378-1393
\bibitem[Horowitz \& Kadau(2009)]{} Horowitz, C.~J., \& Kadau, K.\ 2009, Physical Review Letters, 102, 191102
\bibitem[Jones(1976)]{} Jones, P.~B.\ 1976, \apss, 45, 369 
\bibitem[Jones(2012)]{} Jones, D.~I.\ 2012, \mnras, 420, 2325
\bibitem[Jones et al.(2016)]{} Jones, D.~I., Ashton, G., \& Prix, R.\ 2016, arXiv:1610.03509 
\bibitem[Kramer et al.(2006)]{} Kramer, M., Lyne, A.~G., O'Brien, J.~T., Jordan, C.~A., \& Lorimer, D.~R.\ 2006, Science, 312, 549 
\bibitem[Lander \& Jones(2009)]{} Lander, S.~K., \& Jones, D.~I.\ 2009, \mnras, 395, 2162
\bibitem[Lander(2013)]{} Lander, S.~K.\ 2013, Physical Review Letters, 110, 071101
\bibitem[Lander \& Jones(2017)]{} Lander, S.~K., \& Jones, D.~I.\ 2017, \mnras, 467, 4343 
\bibitem[Lasky \& Glampedakis(2016)]{2016MNRAS.458.1660L} Lasky, P.~D., \& Glampedakis, K.\ 2016, \mnras, 458, 1660
\bibitem[Link(2003)]{} Link, B.\ 2003, Physical Review Letters, 91, 101101 
\bibitem[Link(2006)]{} Link, B.\ 2006, \aap, 458, 881
\bibitem[Lyne et al.(2010)]{} Lyne, A., Hobbs, G., Kramer, M., Stairs, I., \& Stappers, B.\ 2010, Science, 329, 408
\bibitem[Lyne et al.(2013)]{} Lyne, A., Graham-Smith, F., Weltevrede, P., et al.\ 2013, Science, 342, 598
\bibitem[Martynov et al.(2016)]{} Martynov, D.~V., Hall, E.~D., Abbott, B.~P., et al.\ 2016, \prd, 93, 112004
\bibitem[Mestel \& Takhar(1972)]{} Mestel, L., \& Takhar, H.~S.\ 1972, \mnras, 156, 419
%\bibitem[Moore et al.(2015)]{} Moore, C.~J., Cole, R.~H., \& Berry, C.~P.~L.\ 2015, Classical and Quantum Gravity, 32, 015014
\bibitem[Lindblom \& Owen(2002)]{} Lindblom, L., \& Owen, B.~J.\ 2002, \prd, 65, 063006 
\bibitem[Page et al.(2006)]{} Page, D., Geppert, U., \& Weber, F.\ 2006, Nuclear Physics A, 777, 497
\bibitem[Page et al.(2011)]{} Page, D., Prakash, M., Lattimer, J.~M., \& Steiner, A.~W.\ 2011, Physical Review Letters, 106, 081101
\bibitem[Perna \& Gotthelf (2008)]{} Perna, R., Gotthelf, E. V. 2008, \apj, 681, 522
\bibitem[Perna \& Pons (2011)]{} Perna, R., Pons, J. A., 2011, \apjl, 727, 51
\bibitem[Perna et al. (2013)]{} Perna, R., Vigan\'o, D., Pons, J. A., Rea, N. 2013, \mnras, 434, 2362 
\bibitem[Pons \& Perna(2011)]{2} Pons, J.~A., \& Perna, R.\ 2011, \apj, 741, 123 
\bibitem[Philippov et al.(2014)]{} Philippov, A., Tchekhovskoy, A., \& Li, J.~G.\ 2014, \mnras, 441, 1879 
\bibitem[Read et al.(2009)]{} Read, J.~S., Lackey, B.~D., Owen, B.~J., \& Friedman, J.~L.\ 2009, \prd, 79, 124032 
\bibitem[Reisenegger \& Goldreich(1992)]{} Reisenegger, A., \& Goldreich, P.\ 1992, \apj, 395, 240 
\bibitem[Reisenegger(2009)]{} Reisenegger, A.\ 2009, \aap, 499, 557 
\bibitem[Rookyard et al.(2015)]{} Rookyard, S.~C., Weltevrede, P., \& Johnston, S.\ 2015a, \mnras, 446, 3367 
\bibitem[Rookyard et al.(2015)]{} Rookyard, S.~C., Weltevrede, P., \& Johnston, S.\ 2015b, \mnras, 446, 3356 
\bibitem[Sawyer(1989)]{} Sawyer, R.~F.\ 1989, \prd, 39, 3804 
\bibitem[Sedrakian et al.(1999)]{} Sedrakian, A., Wasserman, I., \& Cordes, J.~M.\ 1999, \apj, 524, 341
\bibitem[Shaham(1977)]{} Shaham, J.\ 1977, \apj, 214, 251 
\bibitem[Shternin \& Yakovlev(2008)]{} Shternin, P.~S., \& Yakovlev, D.~G.\ 2008, \prd, 78, 063006
\bibitem[Shternin et al.(2011)]{} Shternin, P.~S., Yakovlev, D.~G., Heinke, C.~O., Ho, W.~C.~G., \& Patnaude, D.~J.\ 2011, \mnras, 412, L108
\bibitem[Stairs et al.(2000)]{} Stairs, I.~H., Lyne, A.~G., \& Shemar, S.~L.\ 2000, \nat, 406, 484
\bibitem[Tauris \& Manchester(1998)]{} Tauris, T.~M., \& Manchester, R.~N.\ 1998, \mnras, 298, 625
\bibitem[Thompson \& Duncan (1995)]{}Thompson C., Duncan R. C., 1995, \mnras, 275, 255
\bibitem[Thompson \& Duncan(2001)]{} Thompson, C., \& Duncan, R.~C.\ 2001, \apj, 561, 980 
\bibitem[Vigan\'o et al. (2013)]{} Vigan\'o, D., Rea, N., Pons, J. A., Perna, R., Aguilera, D. N., Miralles, J. A.
2013, \mnras, 434, 123
\bibitem[Ushomirsky et al.(2000)]{} Ushomirsky, G., Cutler, C., \& Bildsten, L.\ 2000, \mnras, 319, 902 
\bibitem[Watters et al.(2009)]{} Watters, K.~P., Romani, R.~W., Weltevrede, P., \& Johnston, S.\ 2009, \apj, 695, 1289
\bibitem[Weltevrede \& Johnston(2008)]{} Weltevrede, P., \& Johnston, S.\ 2008, \mnras, 387, 1755
\bibitem[Yakovlev et al.(2001)]{} Yakovlev, D.~G., Kaminker, A.~D., Gnedin, O.~Y., \& Haensel, P.\ 2001, \physrep, 354, 1 
\bibitem[Young et al.(2010)]{} Young, M.~D.~T., Chan, L.~S., Burman, R.~R., \& Blair, D.~G.\ 2010, \mnras, 402, 1317
\bibitem[Zanazzi \& Lai(2015)]{} Zanazzi, J.~J., \& Lai, D.\ 2015, \mnras, 451, 695

\end{thebibliography}
\end{document}